\documentclass[a4paper]{article}
\usepackage{amsmath}
\usepackage{amssymb}

\usepackage[dvips]{epsfig}
\addtocounter{secnumdepth}{1}
\newcommand{\dd}{\text{d}}
\newcommand{\ee}{\text{e}}

\newcommand{\ji}{\text{\footnotesize i}}
\newcommand{\p}{\partial}

\newcommand{\beq}{\begin{equation}}
\newcommand{\eeq}{\end{equation}}
\newcommand{\beqa}{\begin{eqnarray}}
\newcommand{\eeqa}{\end{eqnarray}}
\newcommand{\hk}{\hat{\kappa}}

\def\lsim{\:\raisebox{-0.5ex}{$\stackrel{\textstyle<}{\sim}$}\:}
\def\gsim{\:\raisebox{-0.5ex}{$\stackrel{\textstyle>}{\sim}$}\:}

\begin{document}

\thispagestyle{empty}

\large
\title{\bf Symmetry and species segregation\\[2mm]
	   in diffusion-limited  pair annihilation}

\author{H.J.~Hilhorst,$^{1}$ 
	M.J.~Washenberger,$^2$ and U.C.~T\"auber$^2$ \\[3mm]
{\small $^1$ Laboratoire de Physique Th\'eorique, B\^atiment 210,} \\
{\small Universit\'e de Paris-Sud, 91405 Orsay Cedex, France} \\
{\small $^2$ Center for Stochastic Processes in Science and Engineering and
	Department of Physics,} \\
{\small Virginia Polytechnic Institute and State University, Blacksburg,
	Virginia 24061-0435, USA} \\}

\maketitle
\vspace{10mm}

\begin{abstract}
\noindent 

We consider a system of $q$ diffusing particle species $A_1,A_2,\ldots,A_q$ 
that are all equivalent under a symmetry operation. Pairs of particles may 
annihilate according to $A_i+A_j \to 0$ with reaction rates $k_{ij}$ that 
respect the symmetry, and without self-annihilation ($k_{ii}=0$). In spatial
dimensions $d > 2$ mean-field theory predicts that the total particle density 
decays as $\rho(t) \sim t^{-1}$, {\it provided\,} the system remains spatially
uniform. We determine the conditions on the matrix $k$ under which there exists
a critical segregation dimension $d_{\rm seg}$ below which this uniformity 
condition is violated; the symmetry between the species is then locally broken.
We argue that in those cases the density decay slows down to 
$\rho(t) \sim t^{-d/d_{\rm seg}}$ for $2 < d < d_{\rm seg}$. We show that when 
$d_{\rm seg}$ exists, its value can be expressed in terms of the ratio of the 
smallest to the largest eigenvalue of $k$. The existence of a conservation law 
(as in the special two-species annihilation $A+B \to 0$), although sufficient 
for segregation, is shown not to be a {\it necessary} condition for this 
phenomenon to occur. We work out specific examples and present
Monte Carlo simulations compatible with our analytical results. \\

\noindent {\bf PACS \ 05.40.-a, 82.20.-w}
\end{abstract}

\vspace{63mm}

\newpage


\section{Introduction} 
\label{secintroduction}

In nonequilibrium statistical mechanics there exist two annihilation reactions 
that have become prominent model systems. 
These are the pair annihilation processes $A+A \to 0\,$ and $\,A+B \to 0$. 
Interest in fluctuation effects on the first reaction process dates back to the
seminal work by Smoluchowski \cite{SM}, and in more recent times at least to 
Bramson and Griffeath \cite{BG}; and for the second process fluctuations were 
first studied by Ovchinnikov and Zeldovich \cite{OZ}.
The final states of both processes are trivial: All particles (save perhaps a
single one) disappear in the case of $A+A \to 0$, and only the excess number of
particles of the initial majority species is left in the case of $A+B \to 0$. 
The nature of the approach to the final state is, however, nontrivial and has 
been the subject of many investigations.  

The $A+A \to 0$ problem applies, for example, to domain walls in a
one-dimensional system like in the zero-temperature Ising model \cite{DRS}.
Its asymptotic behavior is the same as that of the diffusion-limited 
coagulation process $A+A \to A$ \cite{Pe}. 
The ensuing asymptotic density decay $\sim t^{-1/2}$ has been experimentally 
observed (in an intermediate time window) in the fusion kinetics of 
laser-induced excitons in quasi one-dimensional N(CH$_3$)$_4$MnCl$_3$ (TMMC) 
polymer chains \cite{Kr}.

The study of the two-species annihilation reaction $A+B \to 0$ was motivated 
originally by the cosmological problem of matter-antimatter annihilation. 
Ovchinnikov and Zeldovich \cite{OZ}, and independently a few years later
Toussaint and Wilczek \cite{TW}, asked whether in such a simple annihilation 
model it would be possible that locally in space only, say, matter would be 
left. The answer rapidly turned out to be positive:
The $A+B \to 0$ system exhibits the phenomenon of species ``segregation'', that
is, the emergence of ever growing single-species domains (either $A$ or $B$). 
As a result, after a short initial period, annihilation takes place only in 
``reaction zones'' where the domains border, and the decay of the total density
is slowed down markedly.
Indeed, a non-classical decay $\sim t^{-3/4}$ has been experimentally confirmed
in three dimensions in a calcium / fluorophore reaction system \cite{Mo}.
The condition for segregation to occur is that the spatial dimension $d\,$ be 
less than the critical {\it segregation dimension\,} $d_{\rm seg}$, which in 
the case of $A+B \to 0$ is equal to $d_{\rm seg}=4$. 
Early work on this two-species problem was done by Kang and Redner \cite{KR}; 
clever heuristic reasoning and numerical work on the spatial structure of the 
domains is due to Leyvraz and Redner \cite{LR}; and rigorous results were 
derived by Bramson and Lebowitz \cite{BL}. 

The renormalization group approach to the reaction $A+A \to 0$ was pursued by 
Peliti \cite{Pe} and Lee \cite{L}, and to the process $A+B \to 0$ by Lee and 
Cardy \cite{LC}, and Oerding \cite{Oe}.
The strong interest in these two reactions has sparked off research on many 
related reaction-diffusion problems by various techniques. Here we will
refer only to general introductions and overviews, {\it e.g.} 
Refs.~\cite{KK}-\cite{HTV}.

In 1986 Ben-Avraham and Redner \cite{bAR}, crediting Kang, introduced a 
$q$-species pair annihilation problem that interpolates between these two 
models, and hence puts them in a common perspective. They considered a system 
of $q$ distinct particle species $A_1,A_2,\ldots\!,A_q$, all propagating with 
the same difffusion constant, and reacting according to 
\begin{equation}
   A_i + A_j \to 0 \qquad (1 \leq i < j \leq q)
\label{defMAM}
\end{equation}
with a single fixed rate.
For $q=2$ and $q=\infty$ this $q$-species {\it Mutual Annihilation Model\,} 
($q$-MAM) reduces to the well-understood paradigmatic cases $A+B \to 0$ and
$A+A \to 0$, respectively. (For the latter case, note that the probability of 
identical species to meet becomes vanishingly small in the limit of 
$q \to \infty$; hence any particle encounter will lead to a reaction, and the
distinction of different species becomes meaningless.) 
The general situation has recently been studied analytically and by Monte Carlo
simulations in Refs.~\cite{DHT,ZDbA,HDWT}.

For equal initial densities of all species the asymptotic decay with time $t$ 
of the total particle density $\rho(t)$ follows a power law $\sim t^{-\alpha}$.
In Table I we summarize the values of the decay exponent $\alpha$, known or 
believed to be exact, as a function of $q$ and of the spatial dimension $d$
\cite{DHT,ZDbA,HDWT}.
We also indicate whether or not the density decay is accompanied by particle
species segregation.
The results for $d=1$ and $q=3,4,\ldots$ are based on the exact solution of a
modified model argued by the authors of Refs.~\cite{DHT,HDWT} to be equivalent 
to the one-dimensional $q$-MAM in the large-time limit.
\begin{table}[tbh]
\begin{center}
\renewcommand{\arraystretch}{1.5}
\begin{tabular}{||c|c|c|c||}
\hline
            & $q=2$          & $q=3,4,\ldots$ & $q=\infty$ \\[1mm]
            & $A + B \to 0$  &                & $A + A \to 0$ \\
\hline
 $d\geq 4 $ & $1$            & $1$               & $1$   \\
\cline{1-2}
 $2<d<4$    &                &                   &     \\
\cline{1-1}
\cline{3-4}
 $d=2$      & $d/4 \ (*)$    &\phantom{($\log$)\,} $1$\,\,($\log$)   
                             &\phantom{($\log$)\,} $1$\,\,($\log$) \\
\cline{1-1} \cline{3-4}
 $1<d<2$    &        	     &                   & $d/2$ \\
\hline
 $d=1$      & $1/4 \ (*)$    & $(q-1)/2q \ (*)$  & $1/2$ \\  
\hline
\end{tabular}
\end{center}
\noindent Table I.\,\,
{\small The density exponent $\alpha$ of the asymptotic decay law 
$\rho(t) \sim t^{-\alpha}$ for the total partical density in the $q$-species 
Mutual Annihilation Model, as a function of the number of species $q$ and the 
spatial dimension $d$. An asterisk (*) indicates that segregation occurs and
($\log$) signifies the appearance of logarithmic corrections at the upper 
critical dimension $d_c = 2$.} 
\end{table}

The $q$-species MAM is characterized by an {\it upper critical dimension} 
$d_c = 2$, below which mean-field theory breaks down and renormalization is 
needed. Physically, in dimensions $d \leq 2$ the recurrence properties of
random walkers lead to depletion zones about each surviving particle, and to
particle anticorrelations induced by the annihilation kinetics. A systematic
renormalization group treatment, wherein the spatial dimension $d$ can be
treated as a continuous variable, has only been carried through for the cases
$q=\infty$ (equivalent to the reaction $A+A \to 0$) \cite{Pe}, and for $q=2$ 
(the two-species reaction $A+B \to 0$) \cite{LC,Oe}, which explains the empty 
entries in the center of Table I.

The derivation of the decay laws listed in Table I relies heavily on the full
permutational symmetry of the particle species in the $q$-MAM. It is therefore 
natural to ask what happens to these asymptotic results when the symmetry is
lowered. A case of lower symmetry occurs, for example, when the particle 
species are ordered cyclically and annihilation, with rate $k_1$, is possible 
only between particles of two neighboring species along the cycle.
The reaction constants are then given by
\begin{equation}
  k_{ij} = k_1 \left[ \delta_{(i-j)\,{\rm mod}\,q,1} + 
  \delta_{(i-j)\,{\rm mod}\,q,q-1} \right] \qquad (1\leq i,j\leq q) \ .
\label{kijcam}
\end{equation}
We will refer to this particular system as the $q$-species {\it Cyclic 
Annihilation Model} ($q$-CAM). Other examples that display lower than 
permutational symmetry may easily be constructed.

In this work we develop, within the mean-field framework ({\it i.e.}, in 
dimensions $d>2$), a general method for finding the exponent of the total 
density decay law $\rho(t) \sim t^{-\alpha}$, and determining whether or not 
segregation occurs in a $q$-species model where the reaction rates $k_{ij}$ 
lower the full permutational symmetry. 

In Sec.~\ref{secmodel} we present our model, which is fully defined by the 
$q \times q$ matrix $k$ of annihilation rates. 
In Sec.~\ref{secmeanfield} we study how the particle numbers of the individual 
species fluctuate around their average densities at any given time.
In Sec.~\ref{secfinitevolume} we show, by comparing the average decay law with 
the fluctuations around it, how the reactions come to an end in a finite 
$d$-dimensional volume $L^d$, and we determine the particle densities in the 
final state. 
In Sec.~\ref{secfinitedimension} we treat the full $d$-dimensional space as 
built up from blocks of size $L^d$ that do not communicate for times less than 
the diffusion time, which is of order $L^2$. 
This allows us to derive our general results concerning the occurrence of 
segregation and the exponents in the density decay law. 
In Sec.~\ref{secexamples} we consider a specific  example which, by means of a 
suitable parameter $\lambda$, interpolates between the $q$-CAM (for 
$\lambda=0$) and the $q$-MAM (for $\lambda=1$).
In Sec.~\ref{secsimulations} we present Monte Carlo simulations which, although
preliminary and of limited statistical accuracy, are compatible with our 
analytical findings. 
In Sec.~\ref{secdiffeq} we comment on an alternative approach, namely via rate 
equations with an additional particle diffusion term, before we conclude in 
Sec.~\ref{secconclusion}.

Our considerations allow us to answer another question as well:
In the two-species $A+B \to 0$ system the appearance of segregation is usually
explained heuristically through arguments invoking the local conservation of 
the difference between the number of $A$ and $B$ particles.
This conservation law thus seemed to be at the origin of the segregation 
phenomenon. 
However, the discovery \cite{DHT} that the one-dimensional MAM exhibits 
segregation for all $q<\infty$, even though it is not subject to any 
conservation law, cast doubt upon this apparent direct link between 
conservation laws and segregation. 
This doubt was subsequently 
reenforced by a heuristic argument according to which
the MAM should exhibit segregation in the entire region of the $q$-$d$ plane 
delimited by $1<q<3$ and $2<d<4/(q-1)$ \cite{HDWT}. (The only weakness of this 
argument is that the region concerned by it does not contain any points with 
integer ``physical'' $q$ and $d$, except for $q=2$, where the 
conservation law applies.)
This present work clarifies this issue: We show that the existence of a 
conservation law, although a {\it sufficient\,} condition for segregation in
dimensions $d<4$, does {\it not\,} constitute a {\it necessary\,} condition:
Segregation may in fact occur in the absence of any conservation law.

\section{A general pair annihilation system}
\label{secmodel}

We consider a system of $q$ distinct diffusing particle species 
$A_1, A_2,\ldots,A_q$ subject to the pair annihilation reactions
\begin{equation}
  A_i+A_j \overset{k_{ij}}{\longrightarrow} 0 \qquad (1\leq i\neq j \leq q) \ .
\label{reactioneqns}
\end{equation}
Here the $k_{ij}=k_{ji}$ represent the reaction rates per unit of density; 
they are constrained only by the requirements that they be non-negative and 
that there exist a symmetry operation under which all particle species are 
equivalent. 
We will set $k_{ii}=0$, implying that no self-annihilation is possible. 
Without loss of generality we may consider the matrix $k$ to be irreducible, 
{\it i.e.}, the system of reactants cannot be decomposed into mutually 
noninteracting subsystems.
The $q$-species MAM and the $q$-species CAM discussed in the Introduction are 
obtained as special cases when all reaction rates in Eq.~(\ref{reactioneqns}) 
are equal and when they are given by Eq.~(\ref{kijcam}), respectively.

We shall therefore proceed to analyze the mean-field behavior of the system
(\ref{reactioneqns}) for a general non-negative traceless symmetric matrix $k$.
Our focus will be on the power law for the density decay, 
$\rho(t) \sim t^{-\alpha}$, and on the segregation properties of this system. 
Straightforward scaling analysis tells us that for pair annihilation processes
such as considered here, very generally mean-field theory is applicable in
spatial dimensions $d>d_c=2$, and, likely with logarithmic corrections, at the 
upper critical dimension $d_c=2$. 
We shall operate entirely at the mean-field level, and hence our results will 
concern dimensions $d \geq 2$. 
In our discussion we will briefly touch upon implications for lower dimensions 
$d<2$.

Note that the system of reactions (\ref{reactioneqns}) may be visualized as a 
graph ${\cal G}$ with $q$ vertices representing the $q$ species, in which bonds
carrying weights $k_{ij}$ connect the vertices of pairs of species that may 
react with each other. At certain points in our discussion it will be 
convenient to refer to this graph representation.

\section{Mean-field theory and fluctuations}
\label{secmeanfield}

\subsection{Averages}

Let the stochastic variable $N_i(t)$ denote the particle number of species 
$A_i$ present at time $t$ in a system of volume $L^d$, and let 
$n_i(t) = \langle N_i(t) \rangle / L^d$ be the average density of the $A_i$ for
$i=1,2,\ldots,q$. 
Here $\langle\ldots\rangle$ is an average
with respect to (i) the initial
distribution of the $N_i$ at time $t=0$ and (ii) all realizations of
the stochastic time evolution.
In order to express the mean-field rate equations 
$\dd n_i/\dd t = - \sum_j{k}_{ij} \, n_i n_j$ in terms of dimensionless 
variables, we set $k_0 = \sum_j k_{ij}$ (which because of the symmetry between
all species is independent of $i$) and $n_0=q^{-1} \sum_i n_i(0)$. 
We furthermore define the quantities $\rho_i(t) = n_i(t) / n_0$, 
$\kappa_{ij} = k_{ij} / k_0$, and $\tau = n_0 k_0 \, t$. 
The mean-field equations then become
\begin{equation}
  \frac{\dd \rho_i}{\dd \tau} = - \sum_j \kappa_{ij} \, \rho_i \rho_j \ ,
\label{meanfieldeqns}
\end{equation}
now with the normalization $\sum_j \kappa_{ij} = 1$.
If all initial densities $n_i(0) = n_0$ and hence $\rho_i(0) = 1$ for all $i$, 
then the solution of Eq.~(\ref{meanfieldeqns}) reads
\begin{equation}
  \rho_i(\tau) = \rho(\tau) = \frac{1}{1+\tau} \ ,
\label{mfaverage}
\end{equation}
which tends to zero as a power law with decay exponent $\alpha = 1$. 
For unequal initial densities the asymptotic decay will generically be 
exponential (readily generalized to stretched exponential for $d \leq 2$ as a 
consequence of the reaction rate renormalization \cite{KR,BL,HTV}), with one or
several of the species tending toward a positive limit density 
$\rho_i(\infty)$; this case will not be considered any further below.

\subsection{Fluctuations}

In principle, the fluctuations around this mean-field average are encoded in 
the master equation for the probability distribution $P(N_1,N_2,\ldots,N_q;t)$,
\begin{equation}
  \frac{\dd P}{\dd t} = \! \sum_{1\leq i<j\leq q} \!\! \frac{k_{ij}}{L^d} \, 
  \big[ (N_i+1)(N_j+1) \, P(\ldots,N_i+1,N_j+1,\ldots) - 
   N_i N_j \, P(\ldots,N_i,N_j\ldots) \big] \ .
\label{mastereq}
\end{equation}
A convenient means to extract them proceeds via van Kampen's $\Omega$-expansion
\cite{vK}, as was demonstrated for the $q$-MAM by Ben-Avraham and Redner 
\cite{bAR}, and more recently for a cyclic three-species trapping reaction by 
Ben-Naim and Krapivsky \cite{bNK}, and for a zero-dimensional population 
dynamics model by Newman and McKane \cite{NmK}.

Anticipating that the fluctuations should be of the order of the square root 
of the total particle number $\sum_{i=1}^q N_i$, we set
\begin{equation}
  N_i(t) = n_0 L^d \rho(\tau) + (n_0 L^d)^{1/2} \, \gamma_i(\tau) \ ,
\label{defalphaitau}
\end{equation}
where, due to previous definitions, $\langle \gamma_i(0) \rangle = 0$, 
and transform the probability distribution $P$ on the extensive variables $N_i$
into an equivalent one, to be called $F$, on the intensive variables 
$\gamma_i$, according to
\begin{equation}
  P(N_1,N_2,\ldots,N_q;t) = (n_0L^d)^{q/2} \, 
  F(\gamma_1,\gamma_2,\ldots,\gamma_q;\tau) \ .
\label{defFalpha}
\end{equation}
Expanding the master equation (\ref{mastereq}) for $P$ to second order in 
powers of $(n_0 L^d)^{-1/2}$, as in Refs.\,\cite{bAR}, \cite{bNK}, and 
\cite{NmK}, and exploiting the rate equations (\ref{meanfieldeqns}) for
$\rho_i(\tau) = \rho(\tau)$ then yields a Fokker-Planck equation with 
time-dependent coefficients,
\begin{equation}
  \partial_\tau F = \sum_{1\leq i<j\leq q} \kappa_{ij} 
  \big[ \rho (\partial_i + \partial_j) (\gamma_i + \gamma_{j}) \,
  + \, \tfrac{1}{2} \rho^2 (\partial_i + \partial_j)^2 \big] F \ ,
\label{FPeqn}
\end{equation}
where $\partial_i \equiv \partial/\partial\gamma_i$. 
This equation is valid provided the second term on the r.h.s. of 
Eq.~(\ref{defalphaitau}) remains much smaller than the first one.

\subsection{Equations for the second moments}
\label{sec2secondmoments}

From the Fokker-Planck equation (\ref{FPeqn}) we may readily deduce 
equations for the time evolution of the averages $\langle \gamma_i \rangle$, 
the variances $\langle \gamma^2_i \rangle \equiv \Gamma_{ii}$, and the 
covariances $\langle \gamma_i \gamma_j \rangle \equiv \Gamma_{ij}$. 
Since $\dd \langle \gamma_i \rangle / \dd \tau = - \rho 
\langle \gamma_i \rangle - \rho \sum_j \kappa_{ij} \langle \gamma_j \rangle$, 
and given that the $\langle\gamma_i\rangle$ are zero initially,
we see that they vanish for all times.
The covariance matrix $\Gamma$ satisfies
\begin{equation}
  \frac{\dd \Gamma_{ij}}{\dd\tau} = \rho^2 \, (\kappa_{ij} + \delta_{ij})
  - 2 \rho \, \Gamma_{ij} - \rho \sum_\ell \kappa_{i\ell} \, \Gamma_{j\ell} 
  - \rho \sum_\ell \kappa_{j\ell} \, \Gamma_{i\ell} \ .
\label{eqnsGlmt}
\end{equation}
Now let $U$ be the real unitary matrix with elements $U_{\mu i}$ that renders
$\hat{\kappa} = U \kappa U^{-1}$ diagonal, and denote its eigenvalues by 
$\hat{\kappa}_\mu$.
In the same manner, we define $\hat \Gamma = U \Gamma U^{-1}$.
Upon transforming Eq.~(\ref{eqnsGlmt}) to these new variables we obtain
\begin{equation}
  \frac{\dd \hat\Gamma_{\mu\nu}}{\dd\tau} = 
  \big[ 1 + \tfrac{1}{2} (\hat{\kappa}_\mu + \hat{\kappa}_\nu) \big] \, 
  (\rho^2 \, \delta_{\mu\nu} - 2 \rho \, \hat\Gamma_{\mu\nu}) \ .
\label{eqnGmunut}
\end{equation}
Let us suppose that the particle numbers initially have independent and 
identical fluctuations, {\it i.e.,} 
$\Gamma_{ij}(0) = \Gamma_0 \, \delta_{ij}$. 
Then also $\hat\Gamma_{\mu\nu}(0) = \Gamma_0 \, \delta_{\mu\nu}$, and it 
follows from Eq.\,(\ref{eqnGmunut}) that $\hat\Gamma_{\mu\nu}(\tau) = 0$ for 
$\mu \not= \nu$ at all $\tau > 0$.
The remaining equation of interest is
\begin{equation}
  \frac{\dd \hat\Gamma_{\mu\mu}}{\dd\tau} = (1+\hat{\kappa}_\mu) \rho^2 
  - 2 (1+\hat{\kappa}_\mu) \rho \, \hat\Gamma_{\mu\mu} \ .
\label{eqnGmumutau}
\end{equation}
Notice that since $\rho$ depends on time, it is not useful to scale the time 
$\tau$ in (\ref{eqnGmumutau}) with the factor $1+\hat{\kappa}_\mu$. 
The normalization relation $\sum_j \kappa_{ij} = 1$ implies that 
$\hat{\kappa}_0 \equiv 1$ is an eigenvalue of $\kappa$ with eigenvector 
$(1,1,\ldots,1)$. 
By the Perron-Frobenius theorem, and since the $\hat{\kappa}_\mu$ are 
necessarily real, we have $-1 \leq \hat{\kappa}_{\mu} \leq 1$ for all $\mu$.
In case $\hat{\kappa}_\mu = -1$, the r.h.s. of Eq.\,(\ref{eqnGmumutau}) 
vanishes and the initial fluctuation $\hat\Gamma_{\mu\mu}$ does not decay, 
{\it i.e.,} an eigenvalue $-1$ signals the presence of a conservation law.

\subsection{Solution of the moment equations}

The solution of Eq.~(\ref{eqnGmumutau}), except when 
$\hat{\kappa}_\mu = -\tfrac{1}{2}$, is a sum of two power laws,
\begin{equation}
  \hat\Gamma_{\mu\mu}(\tau) = \big( \Gamma_0 - K_\mu \big) 
  (1+\tau)^{-2 (1+\hat{\kappa}_\mu)} \, + \, K_\mu (1+\tau)^{-1} \ ,
\label{solnflucts}
\end{equation}
where we have introduced the coefficient 
$K_\mu = (1+\hat{\kappa}_\mu) / (1+2\hat{\kappa}_\mu)$.
Note that the amplitude of (only) the first term in Eq.\,(\ref{solnflucts}) 
depends on the initial fluctuation strength $\Gamma_0$. 
Its exponent $2 (1+\hat{\kappa}_\mu)$ varies for the different modes $\mu$, 
which shows that this problem is characterized by a spectrum of power laws.
The second term in Eq.~(\ref{solnflucts}) is just proportional to the density 
$\rho(\tau)$ and hence decays as $\tau^{-1}$. 
Therefore, if $\hat{\kappa}_\mu < - \tfrac{1}{2}$, the first term decays more 
slowly than the second one (and vice versa for 
$\hat{\kappa}_\mu > - \tfrac{1}{2}$).
When it so happens that $\kappa_\mu = - \tfrac{1}{2}$, then
Eq.~(\ref{solnflucts}) should be
replaced with the special logarithmic solution 
\begin{equation}
  \hat{\Gamma}_{\mu\mu}(\tau) = \Gamma_0 (1+\tau)^{-1} \, + \,
  \tfrac{1}{2} (1+\tau)^{-1} \log(1+\tau) \ .
\label{solnfluctslog}
\end{equation}
The variance of the particle number for species $i$ is given by 
\begin{equation}
  \langle \Delta N_i^2(t) \rangle = q^{-1} \sum_i \langle \Delta N_i^2(t) 
  \rangle = n_0 L^d q^{-1} \sum_i \Gamma_{ii}(\tau) =
  n_0 L^d q^{-1} \sum_\mu \hat\Gamma_{\mu\mu}(\tau) \ .
\label{expravvar}
\end{equation}
Let us denote the algebraically smallest eigenvalue of $\kappa$ by 
$\hat{\kappa}_{\rm m}$, its degeneracy by $c_{\rm m}$, and the corresponding 
value of the coefficient $K_\mu$ by $K_{\rm m}$. 
For asymptotically large times we then deduce from 
Eqs.~(\ref{solnflucts})-(\ref{expravvar}) the behavior
\begin{equation}
  \Gamma_{ii}(\tau) \simeq \left\{ \begin{array}{ll} K \, \tau^{-1} &\\[2mm]
  \tfrac{1}{2} \tau^{-1} \log\tau &\\[2mm]
  c_{\rm m} \, q^{-1} (\Gamma_0-K_{\rm m}) \, 
  \tau^{-2(1+\hat{\kappa}_{\rm m})} & \end{array} \right.
  \quad \Big( \hat{\kappa}_{\rm m} \gtreqqless -\tfrac{1}{2} \, , \quad
  \tau \to \infty \Big) \ ,
\label{fluctsllaspt}
\end{equation}
with the abbreviation $K = q^{-1} \sum_\mu K_\mu$.

\section{Final state in a finite volume}
\label{secfinitevolume}

We are now ready to exploit the results of the preceding sections.
We begin by posing the question whether there exists a characteristic time 
$\tau_*(L) = n_0 k_0 t_*(L)$ at which the particle number $N_i$ in the volume 
$L^d$ displays a root mean-square deviation from the average that is of the 
order of the average itself, {\it i.e.,}
\begin{equation}
  \langle N_i(t_*) \rangle \simeq \langle \Delta N_i^2(t_*) \rangle^{1/2} \ .
\label{relavvar}
\end{equation}
Yet the Fokker-Planck equation (\ref{FPeqn}) was derived under the hypothesis 
of small fluctuations (the second term in Eq.\,(\ref{defalphaitau}) was 
supposed to small compared to the first one), and hence ceases to be valid for 
$\tau \gsim \tau_*(L)$.
The issue then is what happens at and beyond this time scale.
Fluctuations satisfying (\ref{relavvar}) indicate that the different particles 
species have widely varying numbers. 
As we argued in the Introduction, in this situation the annihilation 
processes come to an end epxonentially, on the prevailing time scale, through 
the successive extinction of one or several particle species; this final decay
is no longer described by the Fokker-Planck equation (\ref{FPeqn}). 
In the finite volume $L^d$, the final state therefore consists of a collection 
of particles that are not subject to pair annihilations anymore. 
If all $\kappa_{ij} > 0$ for $i \neq j$, then at most a single species can 
survive. 
But if some of the rates $\kappa_{ij}$ vanish, a broader variety of final 
states is possible (we shall return to this point in Sec.~\ref{secexamples}). 
In any of these cases, whereas initially all species were equivalent, the final
state has this symmetry broken.
In fact, this symmetry breaking may be traced back to the first term in 
Eq.~(\ref{solnflucts}), which includes the amplitude $\Gamma_0 - K_\mu$. 
Thus, this symmetry breaking is enhanced by initial fluctuations, as 
represented by $\Gamma_0$, yet it actually {\it persists even for the case
$\Gamma_0 = 0$, {\rm i.e.} if we initially take all particle numbers $N_i(0)$
to be strictly equal}.
We shall discuss several examples of such broken symmetry states in 
Sec.~\ref{secexamples}. 

We turn now to the determination of the characteristic time $\tau_*(L)$.
Utilizing in Eq.~(\ref{relavvar}) that  
$\langle N_i(t) \rangle \simeq n_0 L^d \tau^{-1}$ for $\tau \to \infty$ as well
as inserting the asymptotic results of Eqs.~(\ref{expravvar}) and 
(\ref{fluctsllaspt}), and then solving for $\tau_*(L)$, we find
\begin{equation}
  \tau_*(L) \simeq \left\{ \begin{array}{ll} K^{-1} \, n_0 L^d &\\[2mm]
  2 n_0 L^d / \log (n_0 L^d)  &\\[2mm]
  C^{-1}( n_0 L^d )^{1/|2\hat{\kappa}_{\rm m}|} & \end{array} \right.
  \quad \Big( \hat{\kappa}_{\rm m} \gtreqqless -\tfrac{1}{2} \, , \quad 
  \tau \to \infty \Big) \ ,
\label{solnt*}
\end{equation}
where 
$C = [c_{\rm m} \, q^{-1} (\Gamma_0-K_{\rm m})]^{1/|2\hat{\kappa}_{\rm m}|}$.
The order of magnitude $N_*(L)$ of the particle number for a surviving species 
will be 
$N_*(L) \sim n_0 L^d \rho(\tau_*(L)) \simeq n_0 L^d / \tau_*(L)$, 
and hence is given by
\begin{equation}
  N_*(L) \sim \left\{ \begin{array}{ll} K &\\[2mm]
  \tfrac{1}{2} \log(n_0 L^d) &\\[2mm] 
  C (n_0 L^d)^{1-1/|2 \hat{\kappa}_{\rm m}|} & \end{array} \right. 
  \quad \Big( \hat{\kappa}_{\rm m} \gtreqqless -\tfrac{1}{2} \, , \quad 
  \tau \to \infty \Big) \ .
\label{exprrhot*}
\end{equation}
When $\hat{\kappa}_{\rm m} > -\tfrac{1}{2}$
this surviving number is of order unity, but 
when $\hat{\kappa}_{\rm m} = -\tfrac{1}{2}$
it grows logarithmically with the
volume; and when $\hat{\kappa}_{\rm m} < -\tfrac{1}{2}$
it grows with a positive power of $L$.

\section{Segregation in an infinite volume}
\label{secfinitedimension}

We now consider an infinite volume in which the $q$ annihilating species 
propagate with (uniform) diffusion constant $D$, and apply a heuristic 
analysis:
We imagine this infinite volume divided into hypercubic subvolumes of size 
$L^d$.
For times less than the characteristic diffusion time, {\it i.e.,} for 
$\tau \lsim \tau_{\rm diff}(L)\sim L^2 / D$, the subvolumes can be treated as 
effectively independent, and
hence the results of Eqs.~(\ref{solnt*}) and
(\ref{exprrhot*}) apply to each of them.
We will discuss the four distinct cases separately.

(i) {\it Case $\, -\tfrac{1}{2} < \hat{\kappa}_{\rm m} \leq 1$}.\
From Eq.~(\ref{exprrhot*}) we infer in this situation that in an isolated 
volume the particle number fluctuations become of the order of the average 
itself only at times when the total particle numbers $N_*(L) \sim K$ have 
decreased to order unity themselves. 
In any case, in dimensions $d > 2$ the time scale at which this happens, 
$\tau_*(L) \sim L^d$, is much larger than the diffusion time 
$\tau_{\rm diff}(L) \sim L^2$. 
Hence in an infinite system the subvolumes begin to mix diffusively long
before the fluctuations of their particle numbers
become of the order of the averages. 
This is tantamount to stating that for 
$\hat{\kappa}_{\rm {m}} > -\tfrac{1}{2}$ there is no particle segregation.

(ii) {\it Case $\, -1 < \hat{\kappa}_{\rm m} < -\tfrac{1}{2}$}.\
In this case Eq.~(\ref{solnt*}) states that the fluctuations become of the 
order of the average value $\rho(\tau)$ at a characteristic time 
$\tau_*(L) \sim L^{d / |2 \hat{\kappa}_{\rm m}|}$, and in an isolated volume 
the annihilation processes will come to an end on this same time scale.
Similarly in an infinite system, subject to the condition that
\begin{equation}
  \tau_*(L) \ll \tau_{\rm diff}(L) \ ,
\label{reltimes}
\end{equation} 
each subvolume of size $L^d$ will reach a quasi-final state, to be referred to 
as a ``domain'', and further reactions are possible only at the ``reaction 
zones'' separating the domains.
The quasi-final state in each subvolume is reached independently of its 
neighbors, and the emergence of these different disjoint quasi-final states 
constitutes the phenomenon of segregation into domains.
The condition (\ref{reltimes}) implies that segregation occurs in dimensions 
below a {\it critical segregation dimension} $d_{\rm seg}\,$
\begin{equation}
  d_{\rm seg} = 4 \, |\hat \kappa_{\rm m}| \ .
\label{exprdseg}
\end{equation}
It therefore appears that in the present case the critical segregation 
dimension for diffusion-limited multi-species pair annihilation processes is 
always constrained by the interval $2 < d_{\rm seg} < 4$.
The dimensionless particle density in a given domain will be of the order
$\rho_*(L) = N_*(L) / (n_0 L^d)$ and cannot decrease further until diffusion 
between neighboring domains permits new reactions to take place. 
Hence we conclude that at any time $\tau$ the particle density in an infinite 
system equals $\rho_*(L_{\rm diff}(t))$, where 
$L_{\rm diff}(t) \sim (Dt)^{1/2} = (D \tau / n_0 k_0)^{1/2}$.
By combining these relations we find that the density decays asymptotically as
\begin{equation}
  \rho(\tau) \sim (C' \tau)^{-d/d_{\rm seg}} \qquad (2 < d <d_{\rm seg}) \ ,
\label{densdecay}
\end{equation}
where 
$C' = [c_{\rm m} \, q^{-1} (\Gamma_0-K_{\rm m})]^{-2/d} n_0^{2/d-1} D / k_0$ is
a dimensionless constant. 
Thus we obtain $\alpha \equiv d/d_{\rm seg}$ for $2 < d < d_{\rm seg}$. 

(iii) {\it Case $\, \hat{\kappa}_{\rm m} = - \tfrac{1}{2}$}.\
In this case $d_{\rm seg} = 2$, which is also the upper critical dimension 
$d_c$ where mean-field theory is only marginally applicable. 
Nevertheless, upon following the same reasoning as above we find the asymptotic
density decay law
\begin{equation}
  \rho(\tau) \sim (k_0/2D) \, \tau^{-1} \log (D \tau / k_0) \qquad 
  (d = d_{\rm seg} = d_c = 2)
\label{rhotaudseg}
\end{equation}
in two dimensions.
Based on renormalization group arguments and Monte Carlo simulations, this 
result was predicted in particular for the $3$-MAM \cite{DHT,ZDbA,HDWT}, for 
which indeed $\hk_{\rm m} = -\tfrac{1}{2}$ (see Subsec.~\ref{subsecqmam} 
below).
It is remarkable that the logarithmic corrections (\ref{rhotaudseg}) take the
same form as quite generically predicted for pair annihilation reactions in
two dimensions as consequence of the required reaction rate renormalizations
induced by the appearance of depletion zones in low dimensions.
Although segregation effects are very difficult to capture within the framework
of renormalized field theory (compare the discussion in Ref.~\cite{HDWT}), we
therefore hypothesize, but cannot prove, that no additional logarithmic factors
beyond those exhibited in Eq.~(\ref{rhotaudseg}) appear in the general case.

(iv) {\it Case $\, \hat{\kappa}_{\rm m} = -1$}.\
This special case, for which $d_{\rm seg} = 4$, occurs when the graph 
${\cal G}$ is bipartite, {\it i.e.}, when it is comprised of two subsets of 
vertices such that all bonds are between vertices of different subsets.
Since these subsets are necessarily equivalent under symmetry, this obviously
requires that $q$ be even, and we may relabel the sets to
$\{1,2,\ldots,q/2\}$ 
and $\{q/2+1,q/2+2,\ldots,q\}$.
The particle numbers then obey the {\em local conservation law}
\begin{equation}
  \sum_{i=1}^{q/2} N_i(t) \, - \sum_{i=q/2+1}^q N_i(t) \, = \, \mbox{constant}
  \ ,
\end{equation} 
and it is readily verified that $\kappa$ has an eigenvector 
$(1,\ldots,1,-1,\ldots,-1)$ with eigenvalue $-1$. 
For $\hat{\kappa}_{\rm {m}} = -1$ we moreover infer from Eq.~(\ref{solnflucts})
that $\hat\Gamma_{\mu\mu}(\tau) = \Gamma_0$, which establishes that as a direct
consequence of the conservation law the initial fluctuations in the
total particle numbers of the two subsets do not relax.
Examples where this happens are the $2$-MAM (the two-species pair annihilation
reaction $A+B \to 0$) and the $q$-CAM with even $q$ (see 
Subsec.~\ref{subsecqcam}).

\section{An explicit example}
\label{secexamples}

We proceed to investigate a specific $q$-species pair annihilation system that 
depends on a parameter $\lambda$, interpolating for $0 \leq \lambda \leq 1$ 
between the $q$-CAM and the $q$-MAM that were defined in the Introduction.

We imagine the $q$ species arranged in a cycle.
The annihilation rate is set equal to $k_1$ for any pair of species that are 
nearest neighbors along the cycle, and to $k_2$ for any other pair. 
Then the ratio $\lambda \equiv k_2/k_1$ is the only intervening parameter in 
the normalized matrix $\kappa$.
Using the abbreviation ${\cal N}_{q\lambda} = 1 + \frac{1}{2} (q-3) \lambda$, 
one has
\begin{equation}
  \kappa_{ij} \, = \,\big[ \lambda \, + \, (1-\lambda) \,
  \big(\, \delta_{(i-j)\,{\rm mod}\,q,1} + \delta_{(i-j)\,{\rm mod}\,q,q-1}\,
  \big) \big] / \, 2 {\cal N}_{q\lambda} \qquad (i \neq j)
\label{defkappaijgeneral}
\end{equation}
and $\kappa_{ii} = 0$.
For $0 \leq \lambda <1$ this model 
is symmetric only under cyclic translation of 
the species, and the special case $\lambda = 0$ yields the $q$-CAM.
For $\lambda=1$ the model displays full permutation symmetry and we recover 
the $q$-MAM. 
 
It is straightforward to diagonalize $\kappa$ by Fourier transforming in the 
variable $i-j$, and hence the mode labels $\mu, \nu,\ldots$ belong to
the set of  
wavenumbers $p \in \{ 2\pi n/q \}$ with $n = 0,1,\ldots,q-1$.
The corresponding eigenvalues are
\begin{equation}
  \hat{\kappa}_p = q^{-1} \sum_{i,j} \kappa_{ij} \ee^{\ji p (i-j)} = 
  \big[ \tfrac{1}{2} \lambda (q\,\delta_{p0}-1) \,+\, (1-\lambda) \cos p \big] 
  / \,{\cal N}_{q\lambda} \ .
\label{evkappapgeneral}
\end{equation} 
We shall first discuss the two limiting models, the MAM and the CAM, whose
segregation properties turn out to be quite different. 
Next, our study as a function of $\lambda$ will characterize the crossover
between these two cases, thereby shedding light on the nature of the 
segregation mechanism.

\subsection{Mutual Annihilation Model (MAM)}
\label{subsecqmam}

The parameter value $\lambda=1$ defines the Mutual Annihilation Model 
($q$-MAM).
In this case the graph ${\cal G}$ with $q$ vertices is fully connected and 
every bond carries the same weight.
This model, introduced in Ref.~\cite{bAR}, has been extensively studied in
Refs.~\cite{DHT,ZDbA,HDWT}. 
We may reproduce the known results in dimensions $d \geq 2$ as follows.
Eq.~(\ref{evkappapgeneral}) shows that for $\lambda=1$ the matrix $\kappa$ has 
a single eigenvalue $1$ (namely for the mode with $p=0$) and a $(q-1)$-fold
degenerate eigenvalue $\hat{\kappa}_{\rm m}=-1/(q-1)$ (which occurs for all 
$p \neq 0$).
In the two-species model with $q=2$, this latter eigenvalue equals $-1$, whence
we encounter the special case (iv) of Sec.~\ref{secfinitedimension}. 
There is a conservation law, we find $d_{\rm seg}=4$, and for $d<d_{\rm seg}$ 
the density decays as $\rho(\tau) \sim \tau^{-d/4}$ \cite{KR,BL,LC,HDWT}. 

For $2 < q < 3$, the eigenvalue $\hat{\kappa}_{\rm m} = - 1/(q-1)$ lies in the 
interval $[-1,-\frac{1}{2}]$, whence we are confronted with case (ii) of 
Sec.~\ref{secfinitedimension}. 
According to  Eq.~(\ref{exprdseg}), there then
exists a $q$-dependent segregation dimension
\begin{equation}
  d_{\rm seg}(q) = 4 / (q-1) \ , 
\end{equation}
a relation first derived in Ref.~\cite{HDWT}.
Although this case does not encompass any integer values of $q$, it does 
suggest that segregation may occur even in the absence of a conservation law. 
Such a scenario will indeed be confirmed below.

The marginal value $q=3$ corresponds to the special case (iii) of 
Sec.~\ref{secfinitedimension}. 
Indeed, it was concluded in Refs.~\cite{DHT,HDWT} on the basis of 
renormalization group arguments that in two dimensions the total density in
the $3$-MAM decays as $\rho(\tau) \sim \tau^{-1} \log\tau$, in agreement with
Eq.~(\ref{rhotaudseg}).
For $q > 3$ we obtain $d_{\rm seg}(q) < 2$, and our present theory is not
applicable anymore.
The analytical and numerical studies reported in Refs.~\cite{DHT,ZDbA,HDWT} 
however yield species segregation in one dimension, with a 
$q$-dependent density decay exponent $\alpha(q) = (q-1)/2q$.

\subsection{Cyclic Annihilation Model (CAM)}
\label{subsecqcam}

For $\lambda=0$ the cyclic graph ${\cal G}$ only contains nearest-neighbor 
bonds, {\it i.e.}, a particle of a given species can annihilate only with 
particles of one of its two neighboring species.
This model has been called the {\it Cyclic Annihilation Model\,} or $q$-CAM. 
In this case the eigenvalues given in Eq.~(\ref{evkappapgeneral}) reduce to 
$\hat{\kappa}_p = \cos p$.
For {\it even\,} $q$, the mode with $p = \pi$ yields the smallest eigenvalue 
$\hat{\kappa}_{\rm m} = -1$.
But for {\it odd} $q$ the smallest eigenvalue is acquired for 
$p = \pi (1 \pm q^{-1})$, namely $\hat{\kappa}_{\rm m} = - \cos\frac{\pi}{q}$.
It then follows from Eq.~(\ref{exprdseg}) that  
\begin{equation}
  d_{\rm seg} = \left\{ \begin{array}{ll} 4 & \qquad (q=2,4,6,\ldots) \\[2mm]
  4\cos\frac{\pi}{q} & \qquad (q=3,5,7,\ldots) \end{array} \right. \ .
\label{dsegCAM}
\end{equation}
For $q = 2,3,4$ these results for $d_{\rm seg}(q)$ were previously known, since
in these cases the CAM actually coincides with a MAM: the $3$-CAM coincides
with the $3$-MAM, and the $2$-CAM and $4$-CAM are both equivalent to the 
$2$-MAM \cite{DHT,HDWT}.
The first novel case is therefore $q=5$, for which Eq.~(\ref{dsegCAM}) gives
\begin{equation}
  d_{\rm seg}(5) = 1 + \sqrt{5} = 3.236... \ .
\label{eqndseg5}
\end{equation}
It follows that the $5$-CAM exhibits segregation in spatial dimensions $d=3$ 
and $d=2$, and Eq.~(\ref{densdecay}) implies 
that $\rho(\tau) \sim \tau^{-\alpha}$ with
\begin{equation}
  \alpha=\left\{ \begin{array}{ll} \tfrac{1}{2}(\sqrt{5}-1)=0.618... \qquad 
  &(d=2)\\[2mm] \tfrac{3}{4}(\sqrt{5}-1)=0.927... \qquad &(d=3) \end{array}
  \right. \qquad (q=5) \ .
\label{goldenmean}
\end{equation}
(Curiously, the decay exponent in two dimensions is just the golden mean.)
In simulations, aside from the observation of species segregation, direct 
measurement of the exponent values $\alpha$ provides a straightforward means to
verify the present theory (see Sec.~\ref{secsimulations} below).

The domain structure in the case of the $q$-CAM needs to be discussed.
Whereas the MAM is necessarily characterized by single-species domains, this is
no longer true for the CAM.
Let us first consider $q=5$. For the $5$-CAM one can certainly conceive the 
possibility of single-species domains.
Any given particle species, however, does not react with two other species, and
hence is able to coexist with either of those.
It appears evident, therefore, that a single-species domain is not stable 
against penetration by either of the two species with which its particles 
cannot annihilate.
Thus we must suppose that a typical segregated domain will always contain two 
species, say $\ell$ and $\ell'$, out of the five, {\it viz.} any of the 
cyclically equivalent combinations 
$\{\ell\ell'\} = \{13\}, \{24\}, \{35\}, \{41\},$ $\{52\}$. 
These domains are ``stable against penetration'', in the following sense:
Any particle of a different species intruding into 
such a domain would annihilate 
with at least one of the two domain species.

For $q \geq 6$ new questions appear that we will not fully address here.
As for $q=5$, we may list the subsets of particles that can coexist in a domain
where no interactions take place anymore.
There are two subsets with three species, \{135\} and \{246\}, and three 
subsets with only two species, namely \{14\}, \{25\}, and \{36\}.
All these five subsets are ``impenetrable'' to an intruder species.
It is therefore an interesting issue which type of domains are going to 
be formed in the segregation process.

One important conclusion from the $q$-CAM is that for $q=5,7,\ldots$ we 
encounter ``physical'' situations ($q$ and $d$ are integers), where, in spite
of the absence of a conservation law, species segregation into domains emerges.
Note that the segregation mechanism does not even require initial fluctuations 
in the species numbers:
Even if initially absent, such fluctuations are dynamically generated by the 
annihilation reactions, and they decay more slowly than the average particle 
numbers themselves.

\subsection{Interpolating model}

We now allow an arbitrary value for $\lambda$ in the interval $[0,1]$, 
{\it i.e.}, we consider the more general interpolating model.
For even $q$, Eq.~(\ref{evkappapgeneral}) tells us that the eigenvalue 
$\hat{\kappa}_p$ assumes its minimum for $p = \pi$; for odd $q$ one may verify 
that, just as for $\lambda=1$, it becomes minimal for $p$ as close as possible 
to $\pi$, that is, for $p = \pi(1\pm q^{-1})$.
Upon substituting these minimal eigenvalues in Eq.~(\ref{exprdseg}), we obtain
\begin{equation}
  d_{\rm seg}(q) \, = \,\left\{ \begin{array}{ll} 
  (4-2\lambda) / {\cal N}_{q\lambda} & \ q \mbox{ even} \\[2mm]
  \Big[ 2\lambda + 4(1-\lambda) \cos\frac{\pi}{q} \Big] / \, 
  {\cal N}_{q\lambda} & \ q \mbox{ odd} \end{array} \right. \ .
\label{eqndsegqeven}
\end{equation}
For $\lambda = 0,1$, this expression reduces to the limiting values 
found above. It is 
consistent with our mean-field assumptions only as long as it leads to a
$d_{\rm seg}(q)\geq 2$. That is easily seen to imply
that at given $\lambda$, it is meaningful only for
$q$ below a maximum value $q_c(\lambda)$; for $q$ above that value, there is no
segregation in any $d\geq 2$.

\section{Simulations}
\label{secsimulations}

We have performed Monte Carlo simulations in one, two, and three dimensions for
the $q$-CAM with $q = 2,3,4,5$ with the goal to test the theoretical 
predictions for the decay exponent $\alpha$.
For more particle species, our computing resources unfortunately cannot provide
sufficiently reliable data statistics.

The simulation algorithm proceeds as described in Ref.~\cite{HDWT}: 
Starting from an initially random distribution of particles on a 
$d$-dimensional ($d=1,2,3$) hypercubic lattice with periodic boundary 
conditions, the system is evolved as follows: 
A particle is randomly picked.
Next, one of its $2d$ nearest-neighbor sites is selected randomly; if it is 
occupied by a particle of a different species, both particles are removed, 
otherwise the particle hops to the empty site.
The Monte Carlo time is scaled with the total number of particles present
(asynchronous time update).

\begin{figure}
\begin{center}
\includegraphics[scale=0.6]{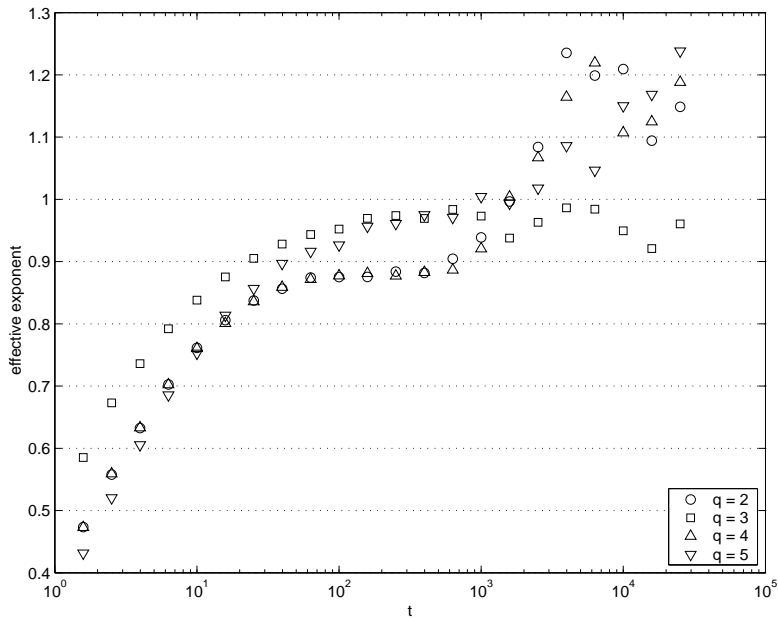}
\caption{\label{3dCAM} Effective decay exponent $\alpha(t)$ for the 
  three-dimensional $q$-CAM with $q=2$, $3$, $4$, and $5$ (random initial 
  conditions, equal initial particle numbers for each species). 
  The data were obtained on a $100 \times 100 \times 100$ cubic lattice, 
  averaged over $20$ runs.}
\end{center}
\end{figure}
Beginning with the data in three dimensions, obtained from simulations on a 
$100 \times 100 \times 100$ cubic lattice by averaging over $20$ runs with 
random initial conditions, we plot the effective decay exponent
\begin{equation}
  \alpha(t) = - \frac{\dd \ln \rho(t)}{\dd \ln t}
\end{equation}
in Fig.~\ref{3dCAM}.
In $d=3$, naturally our statistics is worst, and the data cease to be reliable
at around $t \approx 600$.
The results for even and odd $q$ are clearly distinct; and as anticipated, the 
processes for $q=2$ and $q=4$ are equivalent.
Whereas Eqs.~(\ref{dsegCAM}) and (\ref{densdecay}) predict $\alpha = 3/4$ 
asymptotically for $q$ even, the data show that the effective exponent 
$\alpha(t)$ rather reaches a plateau at $\sim 0.88$.
As discussed in Ref.~\cite{HDWT}, one should however expect the asymptotic
decay law to be somewhat masked by the mean-field behavior $\sim t^{-1}$, which
could explain the combined effective decay exponent closer to $1$.
Indeed, we found the same deviation from the asymptotic value $0.75$ in our
simulations for the $2$-MAM (see Fig.~8 in Ref.~\cite{HDWT}).
The data for odd $q$ are similarly plagued by crossover effects.
For $q=3$, for which the MAM and CAM are equivalent, we see a slow convergence 
towards $\alpha=1$.
The data for $q=5$ might indicate that $\alpha(t)$ reaches a plateau at a 
smaller value, in line with the prediction $\alpha \approx 0.93$ of 
Eq.~(\ref{goldenmean}), but the deteriorating statistics preclude a definite
conclusion.

\begin{figure}
\begin{center}
\includegraphics[scale=0.6]{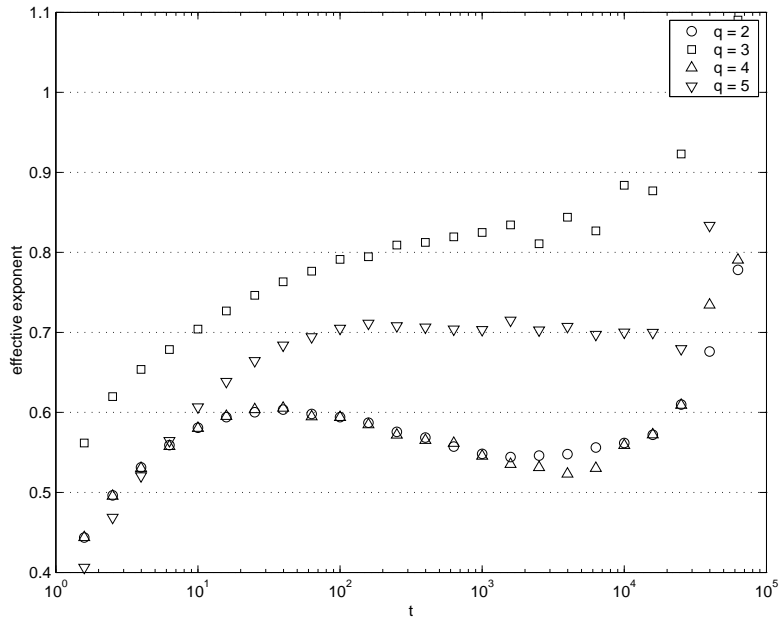}
\caption{\label{2dCAM} Effective decay exponent $\alpha(t)$ for the 
  two-dimensional $q$-CAM with $q=2$, $3$, $4$, and $5$ (random initial 
  conditions, equal initial particle numbers for each species). 
  The data were obtained on a $1000 \times 1000$ square lattice, averaged over 
  $20$ runs.}
\end{center}
\end{figure}
Our two-dimensional results, from $20$ runs on a $1000 \times 1000$ square 
lattice, are depicted in Fig.~\ref{2dCAM}.
The data become unreliable at about $t \approx 1000$.
Once more, the $q$-CAM for $q=2$ and $q=4$ are seen to be equivalent, with
$\alpha(t)$ settling towards the asymptotic value $1/2$, albeit masked again by
the competing mean-field power law, just as for the $2$-MAM (see Fig.~6 in
Ref.~\cite{HDWT}).
Yet now the runs for $q=3$ and $q=5$ yield manifestly different power laws.
In the three-species CAM, the effective exponent is still changing in the time
window accessible to our simulations, running towards $\alpha=1$, perhaps with
logarithmic corrections as predicted by Eq.~(\ref{rhotaudseg}).
For $q=5$, however, we find a plateau value $\sim 0.71$, perhaps with a slowly
decreasing tendency.
This may be interpreted as a combination of the predicted asymptotic decay law 
(\ref{goldenmean}) with the mean-field result $\sim t^{-1}$.

\begin{figure}
\begin{center}
\includegraphics[scale=0.6]{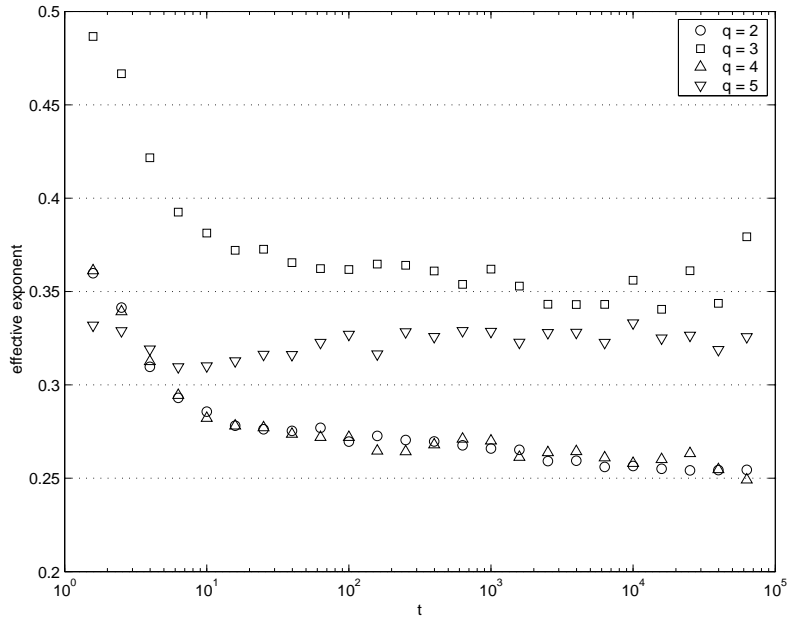}
\caption{\label{1dCAM} Effective decay exponent $\alpha(t)$ for the 
  one-dimensional $q$-CAM with $q=2$, $3$, $4$, and $5$ (random initial 
  conditions, equal initial particle numbers for each species). 
  The data were obtained on a $10^5$ lattice, averaged over $20$ runs.}
\end{center}
\end{figure}
The simulation data in $d=1$, from averaging over again $20$ runs on $10^5$
lattice sites, are reliable at least up to $t \approx 10^4$, and again confirm 
the equivalence of the $q$-CAM with even $q$.
For both $q=2$ and $q=4$ we obtain a slow but definite approach towards the 
predicted $\alpha = 1/4$, as for the $2$-MAM (see Fig.~2 in Ref.~\cite{HDWT}).
For the $3$-CAM, equivalent to the $3$-MAM, we find the expected slow 
convergence towards $\alpha = 1/3$ from above \cite{DHT,ZDbA,HDWT}.
Remarkably, the data for $q=5$, for which no analytical prediction is available
in one dimension, appear to reach $\alpha(t) \approx 0.33$ as well, but faster
and from below.
This is remarkably close to what Eq.~(\ref{densdecay}) would predict if
we applied it (without justification) in dimension $d=1$, namely
$\alpha=1/d_{\rm seg}(5) = 0.309...$, with the
segregation dimension given in Eq.~(\ref{eqndseg5}).

In summary, our simulation results for the $q$-CAM, within the accuracy and
statistical errors of our data, are consistent with the analytical predictions
of the previous sections, if we allow for crossover effects induced by the
presence of the competing mean-field decay law.
The equivalence of the $4$-CAM and $2$-CAM, which are both identical to
the $2$-MAM ($A+B\to 0$), is clearly demonstrated. 
Since the $3$-CAM and $3$-MAM are identical, only our data for $q=5$ yield 
novel results.
Unambiguous identification of the density exponents $\alpha$
would ideally require 
better statistics, that is, simulations on considerably larger systems.

\section{Differential equation approach}
\label{secdiffeq}

Before concluding we wish to make a methodological remark.
One may attempt to address the issue of segregation by considering 
space-dependent particle densities $\rho_i(x,\tau)$ that satisfy the 
reaction-diffusion equations (with diffusivity $D$)
\begin{equation}
  \frac{\p \rho_i(x,\tau)}{\p\tau} \, = \, D \Delta \rho_i
  - \sum_j{\kappa}_{ij} \, \rho_i \rho_j \ .
\label{meanfielddiffeqns}
\end{equation}
For spatially uniform (well-mixed) systems these reduce to the ordinary 
differential equations (\ref{meanfieldeqns}). 
Let us now apply to the system (\ref{meanfielddiffeqns}) what is essentially
a linear stability analysis.

To this end, we write $\rho_i(x,\tau) = \rho(\tau) [1 + \epsilon_i(x,\tau)]$
and linearize with respect to the perturbations $\epsilon_i$. 
With Eqs.~(\ref{meanfieldeqns}) this yields
\begin{equation}
  \frac{\p \epsilon_i(x,\tau)}{\p\tau} \, = \, D \Delta\epsilon_i(x,\tau)
  - \rho(\tau) \sum_j \kappa_{ij} \epsilon_j \ ,
\label{eqnepsi}
\end{equation}
where the neglected nonlinear terms are of order $\rho \, \epsilon^2$.
In terms of the linearly transformed variables
\begin{equation}
  \hat{\epsilon}_\mu(p,\tau) = \sum_j \, U_{\mu j} \int \dd^d x \ 
  \ee^{\ji p\cdot x} \, 
\epsilon_j(x,\tau)
\end{equation}
with $U$ as defined  in Subsec.~\ref{sec2secondmoments}, 
the time evolution equation (\ref{eqnepsi}) becomes diagonal and reads
\begin{equation}
  \frac{\dd \hat{\epsilon}_\mu(p,\tau)}{\dd \tau} = 
  - [D p^2 + \rho(\tau) \hat{\kappa}_\mu\,] \, \epsilon_\mu(p,\tau) \ .
\end{equation}
For given initial $\hat{\epsilon}_\mu(p,0)$ this is solved by
\begin{equation}
  \hat{\epsilon}_\mu(p,\tau) = \hat{\epsilon}_\mu(p,0) \, 
  (1+\tau)^{-\hat{\kappa}_\mu} \, \ee^{-D p^2 \tau} \ .
\label{epsmusol}
\end{equation}
Irrespective of the value of $\hat{\kappa}_\mu$ this solution exists and tends 
to zero as $\tau \to \infty$.
Nevertheless, for $\hat{\kappa}_\mu < - D p^2$ it will increase initially.
In the event that $|\hat{\epsilon}_\mu|$ grows with time such as to be no 
longer negligible with respect to unity at some instant of time, 
this signals that the neglected 
nonlinear terms in the differential equation begin to play a role; in 
particular, they will prevent $\epsilon_i$ from decreasing beyond $-1$ and 
hence $\rho_i$ from turning negative. 
We will argue, however, that in this case the physical justification for the 
differential equations (\ref{meanfielddiffeqns}) breaks down, and that in fact
segregation sets in.

This goal requires going beyond the mathematics and invoking 
the interpretation of the density as an average over discrete particles.
For particles initially randomly and uniformly distributed in a volume $L^d$, 
we typically have that $\hat{\epsilon}_\mu(p,0) \sim L^{-d/2}$.
If this initial fluctuation is negative, and if in the course of time
$\hat{\epsilon}_\mu(p,\tau)$ approaches $-1$, this should be interpreted 
properly as there being an appreciable probability for species extinction.
A necessary
condition for the r.h.s. of Eq.~(\ref{epsmusol}) to become of order unity
reads explicitly $L^{-d/2} (1+\tau)^{-\hat{\kappa}_\mu} \gtrsim 1$, or, since
$\hat{\kappa}_\mu$ must be negative, 
$\tau \gtrsim \tau_{_L} \equiv L^{d / |2 \hat{\kappa}_\mu|}$.
But on these time scales $D p^2 \tau$ must still be much less than unity.
Taking the smallest allowed valued for $p$ in the volume $L^d$, that is,
$p_{_L} \sim 2\pi / L$, the additional condition $D p_{_L} \tau_{_L} \ll 1$ 
yields $L^{d / |2 \hat{\kappa}_\mu| - 2} \ll 1$, which will happen when 
$d < 4|\hat{\kappa}_\mu|$.

Hence the differential equation approach, combined with appropriate
considerations in which the particle discreteness intervenes, 
leads to the identical 
expression for the critical segregation dimension as found in
Eq.~(\ref{exprdseg}) of Sec.~\ref{secfinitedimension}. 
The differential equation approach was followed, essentially, in 
Ref.~\cite{HDWT}.
To our opinion, however, the Fokker-Planck method used in Refs.~\cite{bAR,bNK} 
and in the present work is preferable, since it is based directly
on the more fundamental description of an interacting many-body problem 
in terms of a master equation.
More specifically, the stochasticity taken into account in the differential
equation approach is due only to the random fluctuations present in the initial
state. As a consequence, this approach wrongly
suggests that the initially dominant species
(or, more precisely, their initially dominant mode $\mu$) is also the surviving
one. 

It follows from the Fokker-Planck equation that there is actually an
interplay between initial and dynamically generated
fluctuations, brought out by
the solution (\ref{solnflucts}) of the second moment equations.
This solution contains
a dynamically generated contribution to the fluctuations ({\it viz.}
the terms $\propto K_\mu$), and a contribution due to the initial
fluctuations (the $\Gamma_0$ term).
In systems {\it without} segregation ({\it i.e.}, when 
$\hat{\kappa}_\mu \geq -\frac{1}{2}$) the dynamically generated fluctuations 
become {\it larger} than those due to the initial conditions, which
are eventually forgotten.
In systems {\it with} segregation
(for $\hat{\kappa}_\mu < -\frac{1}{2}$) the initial and dynamical
contributions are of the same order and the initial conditions
co-determine the final state.
In this weaker sense, 
the segregation phenomenon that appears in the cases with
$\hat{\kappa}_\mu < -\frac{1}{2}$ is still linked to the persistence of
initial fluctuations, even in the absence of conservation laws.

\section{Conclusion}
\label{secconclusion}

We have studied a wide class of $q$-species reaction-diffusion systems with 
pair annihilation processes between distinct species. 
This class includes the two well-studied paradigmatic cases $A+A \to 0$ (in the
limit $q \to \infty)$ and $A+B \to 0$ ($q = 2$). Within mean-field theory, 
{\it i.e.} for spatial dimensions $d \geq 2$, we have determined for each 
member of the class (i) whether or not segregation occurs, and (ii) the value 
of the decay exponent $\alpha$ in the asymptotic power law for the total 
particle density. 
Our findings represent a considerable extension of previous work on segregation
in diffusion-limited annihilation reactions. 
Our preliminary 
simulation data are compatible with the analytical results.

Our method builds on ideas that were applied earlier in the context of 
$q$-species models by Ben-Avraham and Redner \cite{bAR}, and more recently also
by Ben-Naim and Krapivsky \cite{bNK} and Newman and McKane \cite{NmK}. 
This approach is obviously not limited to the special cases studied here, but 
we have not aimed at being exhaustive. 
It is straightforward, for example, to drop the restriction of no 
self-annihilation ($k_{ii} = 0$), and redo the analysis.
The method may also be employed to analyze more involved situations, such as 
models with species belonging to two distinct equivalence classes.

Various open problems remain, in particular concerning the nature of the domain
structure in several important cases. 
Furthermore, the theory presented here is unable to address the issue of what 
happens in low dimensions, $d < 2$, where particle anticorrelations become 
manifest and render the mean-field treatment invalid.
We expect that more elaborate simulations will be carried out in the future to 
test our predictions as well as to study questions beyond reach of the present 
theory.

\section*{Acknowledgements}

This material is based upon work supported by the National Science Foundation, 
Division of Materials Research, under Grant No.~DMR-0308548.
The Laboratoire de Physique Th\'eorique of the Universit\'e de Paris-Sud 
is associated with the Centre National de la Recherche Scientifique
as research unit UMR 8627.

\end{document}